\documentstyle[twocolumn,aps]{revtex}
\begin{document}
\input psfig
\draft

%\baselineskip=18pt plus 2pt minus 1pt
%\magnification=1200
%\hsize=5.7truein
%\vsize=8.4truein
%\voffset=24pt
%\hoffset=.1in
%
\twocolumn[\hsize\textwidth\columnwidth\hsize\csname@twocolumnfalse\endcsname
%
% Title Page
%

\title{Optical Conductivity $\sigma(\omega)$ and Resistivity $\rho_{dc}$
\\ of a Hole Doped Spin-Fermion Model for Cuprates}

\author{Mohammad Moraghebi$^1$, Seiji Yunoki$^2$ and Adriana Moreo$^1$}

\address{$^1$Department of Physics, National High Magnetic Field Lab and
MARTECH,\\ Florida State University, Tallahassee, FL 32306, USA}

\address{$^2$Istituto Nazionale di Fisica della Materia and  
International School for
Advanced Studies (SISSA), \\  via Beirut 4, Trieste, Italy}

\date{\today}
\maketitle

\begin{abstract}

The optical conductivity and Drude weight of a Spin-Fermion model for cuprates
are studied as a function of electronic density and temperature.
This model develops stripes and robust D-wave pairing correlations upon hole
doping, and it has the advantage that it can be numerically simulated
without sign problems. Both static and dynamical information can be obtained.
 In this work, it was possible to analyze up to  
$12 \times 12$ site clusters at low
temperatures ranging between $0.01t$ and $0.1t$ 
(between $50K$ and $500K$ for a hopping amplitude $t$$\sim$0.5 eV).
As the temperature is reduced, spectral weight is transferred
from high to low frequencies in agreement with the behavior observed
experimentally. Varying the hole density, the Drude weight has a maximum
at the optimal doping for the model, i.e., at the density where 
the pairing correlations
are stronger. It was also observed that the inverse of the Drude weight,
roughly proportional to the resistivity, decreases $linearly$ with the
temperature at optimal doping, and it is abruptly reduced when robust 
pairing correlations develop upon further reducing the temperature. 
The behavior and general form of the optical conductivity are found to be
in good agreement with experimental results for the cuprates.
Our results also establish the Spin-Fermion model for cuprates as a
reasonable alternative to the $t-J$ model, which is much more difficult to
study accurately.
\end{abstract}

\pacs{PACS numbers: 74.25.Gz, 71.10.Fd}
\vskip2pc]
\narrowtext

\section{Introduction}

Infrared measurements are an important probe of the dynamical properties
of the high critical temperature superconducting cuprates. 
In particular, the real part of the optical conductivity
$\sigma(\omega)$ provides useful insight into the electronic structure of
these materials. \cite{Thomas,Cooper,Uchida}
Measurements of the optical conductivity have been performed in both
hole and electron doped cuprates. Common features observed include
the absence of absorption below the charge transfer gap in the insulating
phase (half-filling). It has also been reported 
the rapid transfer of spectral weight to low
frequencies with increasing hole doping, 
which gives rise to a Drude-like peak at
$\omega=0$ and a broad mid-infrared feature, while the spectral weight
above the charge transfer gap decreases.\cite{Cooper,Uchida} The
reported integrated conductivity up to 4 eV (beyond the charge transfer
gap) in ${\rm La_{2-x}Sr_xCuO_4}$ remains approximately constant with
doping, indicating that spectral weight is redistributed from the charge
transfer band to lower frequencies.\cite{Uchida} 
It is also observed that at low frequencies 
$\sigma(\omega)\approx 1/\omega$ instead of the $1/\omega^2$ behavior
expected in a standard metal.

A number of the above mentioned features are well reproduced by
models of strongly correlated electrons, including the Hubbard, $t-J$,
and related Hamiltonians,\cite{Sega,Fedro} even in regimes
where superconductivity was not numerically detected in the ground state.
This suggests that some of the previously described properties of cuprates
may just be the
effect of strong correlations among the electrons, rather than 
superconductivity which is expected to develop
at much lower
temperatures that those previously studied. 
A problem in the calculation of the optical
conductivity in Hubbard and $t-J$ models is that only small clusters of
about 20 sites can be studied with exact diagonalization 
techniques.\cite{elbio,review}
Alternative approaches, such as the quantum Monte Carlo method 
allows the analysis of larger clusters ($8
\times 8$), but at
very high temperatures and with less precision.\cite{doug}
However, recent investigations have shown that these problems 
can be overcome by using a 
Spin-Fermion model (SFM), which is much easier to study numerically. 
In fact those previous investigations have
reported the presence of
a stable striped ground-state upon doping, and robust
D-wave pairing correlations in the SFM.\cite{we,moham,dwave} The presence of 
phenomenologically observed regimes of the high-$T_c$ cuprates in a simple
model of interacting mobile carriers and localized spins is remarkable,
and opens the way to detailed numerical investigations of relevant physical
quantities. In this context clusters as large as $12\times 12$
can be investigated for low temperatures ranging between 0.01$t$ 
and 0.1$t$, considerably improving upon size and temperature limitations
of the $t-J$ and Hubbard models. In particular,
the effect of the stripes on $\sigma(\omega)$ and its
distribution of spectral weight, and dependence with
temperature can be studied in this case, as reported here.

The organization of the paper is as follows: In section II the
Hamiltonian and the numerical technique are described. Results at
$T\approx 0$ are discussed in section III, while section IV is devoted
to the finite temperature analysis. 
An estimation of the linear dc resistivity
is presented in section V and the conclusions appear in section VI. 

\section{Model and Technique}

The SFM is constructed as an interacting system of
electrons and spins, which mimics phenomenologically the
coexistence of charge and spin degrees of freedom in 
the cuprates.\cite{Pines,Schrieffer,Fedro}. Its Hamiltonian is given by
$$
{\it H=
-t{ \sum_{\langle {\bf ij} \rangle\gamma}(c^{\dagger}_{{\bf i}\gamma}
c_{{\bf j}\gamma}+h.c.)}}
+{\it J
\sum_{{\bf i}}
{\bf{s}}_{\bf i}\cdot{\bf{S}}_{\bf i}
+J'\sum_{\langle {\bf ij} \rangle}{\bf{S}}_{\bf i} \cdot{\bf{S}}_{\bf j}},
\eqno(1)
$$
\noindent where ${\it c^{\dagger}_{{\bf i}\gamma} }$ creates an electron
at site ${\bf i}=({\it i_x,i_y})$ with spin projection $\gamma$,  
${\bf s_i}$=$\it \sum_{\gamma\beta} 
c^{\dagger}_{{\bf i}\gamma}{\bf{\sigma}}_{\gamma\beta}c_{{\bf
i}\beta}$ is the spin of the mobile electron, the  Pauli
matrices are denoted by ${\bf{\sigma}}$,
${\bf{S}_i}$ is the localized
spin at site ${\bf i}$,
${ \langle {\bf ij} \rangle }$ denotes nearest-neighbor (NN)
lattice sites,
${\it t}$ is the NN-hopping amplitude for the electrons,
${\it J}>0$ is an AF coupling between the spins of
the mobile and localized degrees of freedom,
and ${\it J'}>0$ is a direct AF coupling
between the localized spins.
The density $\it \langle n \rangle$=$\it 1-x$ of 
itinerant electrons is controlled by a chemical potential $\mu$. 
Hereafter ${\it t}=1$ will be used as the unit of energy. 
From 
previous phenomenological analysis, 
and as in previous papers \cite{we,Charlie},
the coupling ${\it J}$=2 and the Heisenberg coupling
${\it J'}$=0.05 are selected, since they provide the striped and pairing
states observed in cuprates. 
To simplify the numerical calculations, avoiding the 
sign problem, 
localized spins are assumed to be classical (with $\it |S_{\bf i}|$=1).
This approximation is not as drastic as it appears, and it 
has been extensively discussed in 
detail in previous publications such as 
Ref. \cite{Charlie}. The model will be studied using a
Monte Carlo method, details of which can be found in Ref.~\cite{yuno}.
Periodic boundary conditions (PBC) are used.

The real part of the optical conductivity is calculated as \cite{mahan}
$$
{\it \sigma_{\alpha\alpha}(\omega)={(1-e^{-\beta\omega)}\over{\omega
}} {\pi\over{Z}}\sum_{n,m}e^{-\beta E_n}|\langle
n|J_{\alpha}|m\rangle |^2}
$$
$$
{\it\times\delta(\omega+E_n-E_m)},
\eqno(2)
$$
\noindent where $\alpha=x, y$; Z is the partition function; $|n\rangle$
and $|m \rangle$ denote eigenstates of the system; and $J_{\alpha}$ is
the current operator given by
$$
{\it J_{\alpha}}={{\it it}\over{2}}{\it\sum_{\bf r \alpha \sigma}
(c^{\dagger}_{\bf r+\alpha, \sigma}c_{\bf r, \sigma}-
c^{\dagger}_{\bf r, \sigma}c_{\bf r+\alpha, \sigma})}.
\eqno(3)
$$
\noindent Since we are using PBC, the Drude weight $D_{\alpha}$ is calculated
indirectly, using the two-dimensional sum rule,
$$
\int_0^{\infty} d \omega \sigma_{\alpha\alpha}(\omega)={\pi\over{2N}}\langle
 -K_{\alpha}\rangle,
\eqno(4)
$$
\noindent where $\langle -K_{\alpha}\rangle$ is the average kinetic energy in
the direction $\alpha$ and $N$ is the number of sites in the lattice. 
Assuming the existence of a contribution $D\delta(\omega)$ at zero
energy, we obtain
$$ 
{D_{\alpha}\over{\pi}}={\langle
-K_{\alpha}\rangle\over{2N}}-{1\over{\pi}}\int_{0^+}^{\infty} d \omega
\sigma_{\alpha\alpha}(\omega),
\eqno(5)
$$
\noindent 
This procedure to calculate the Drude weight is common practice and it has
been discussed extensively in reviews.\cite{review}
In this work, both directions $x$ and $y$ must be analyzed independently
since at low
temperature, $T<0.025t$, the symmetry under rotations is broken due to
the formation of stripes. As a consequence, the optical conductivity will be
measured along the direction parallel and perpendicular to those stripes.

In numerical calculations of the optical conductivity it is customary to
replace the delta functions in Eq.(2) by Lorentzians of width
$\epsilon$. The smearing of the deltas helps
to simulate effects not considered in the microscopic
Hamiltonians, such as disorder which contributes to the experimentally observed
broadening of the peaks.\cite{review} 
This peak broadening also occurs naturally in
the Monte Carlo simulations since the position of the poles varies
slightly between successive iterations. 
Here, the incoherent part of the optical
conductivity is calculated by effectively giving a width
$\epsilon\approx 0.01t$ to the
delta functions. That same value of $\epsilon$ will be used in the Lorentzian
for the Drude weight.

\section{Results at $T$$\sim$0}

\begin{figure}[thbp]
\centerline{\psfig{figure=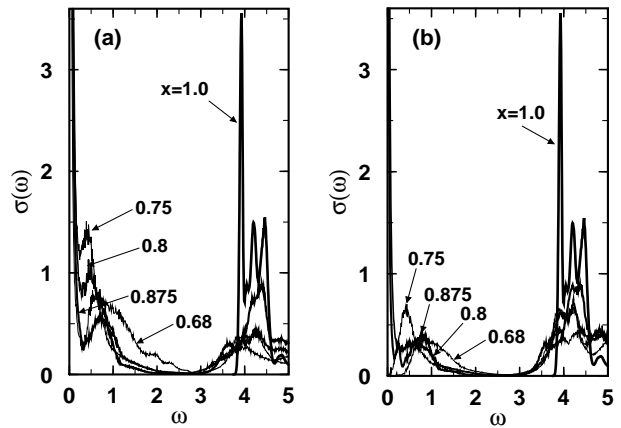,width=8cm}}
\vskip 0.2cm
\caption{(a) The optical conductivity versus $\omega$ at $T=0.01$ in the 
direction
perpendicular to the stripes and different densities on a $12 \times 12$
cluster.
(b) Same as (a) but in the direction parallel to the stripes.}
\end{figure}

In Fig.1a, we present the real part of the optical conductivity measured
in the direction perpendicular to the stripes on a $12 \times 12$ lattice at
$T=0.01t$ for different values of the electronic density $\langle n \rangle$.
At half-filling all the spectral weight appears beyond $\omega\approx
3.8t$. This corresponds to about 2 eV if we assume $t\approx$ 0.5eV, in
agreement with experimental results. This half-filling gap is created
by the coupling $J$, which plays a role analogous to $U$ in the Hubbard model.
In fact, $J$ suppresses double occupancy of the mobile carriers
as strongly as $U$ does. This important point has been extensively discussed
particularly in the manganite literature where a similar model, but with
different signs and values of couplings is used to describe $e_g$ electrons
moving in a $t_{2g}$ background.\cite{manga}

\begin{figure}[thbp]
\centerline{\psfig{figure=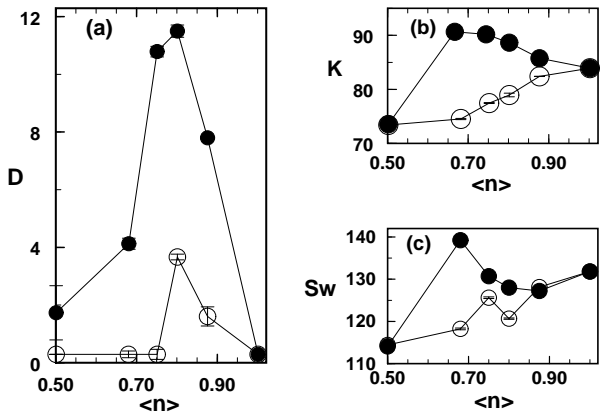,width=8cm}}
\vskip 0.2cm
\caption{(a) The Drude weight as a function of the electronic density 
at $T=0.01$
for currents perpendicular (parallel) to the stripes denoted by filled
(open) circles. (b) The kinetic energy as a function of the electronic
density at $T=0.01$ in the direction perpendicular to the stripes
(filled circles) and parallel to them (open circles). (c) Integral of the
incoherent spectral weight of
$\sigma(\omega)$ versus the electronic density, calculated along 
the direction perpendicular to the
stripes (filled circles) and parallel (open circles).}
\end{figure}

For $\langle n \rangle=0.875$,
spectral weight is observed at low values of $\omega$, again as in the
experimental data for the cuprates. In the same figure, data for 
$\langle n \rangle=0.80$, 0.75 and 0.68 is shown. It can be observed that
spectral weight is transferred from above the gap to low energies in all
cases. At the same time, the Drude weight, $D$, shown in Fig.2a with filled 
circles, becomes
finite with doping, indicating that the system transforms 
from an insulator to a 
conductor. Notice that D reaches its maximum value for
$\langle n \rangle\approx 0.75-0.80$. According to previous investigations,
this corresponds to the optimal doping
for the SFM, i.e., these are the densities at which the D-wave pairing
correlations are the strongest.\cite{dwave} The maximum obtained
in the Drude weight
is due to the interplay between the kinetic energy and the incoherent
spectral weight, which are shown in Fig.2b and c respectively. It is
interesting to notice that in previous studies performed on the large U 
Hubbard and $t-J$ models, it was observed that the kinetic energy reaches a 
maximum
at quarter-filling in regimes without stripes.\cite{tJ} 
However, in the SFM, the kinetic
energy perpendicular to the stripes has a maximum for 
$\langle n \rangle\approx 0.7$ (see Fig.2b). This is in agreement with
experimental results which found that the plasma frequency, proportional
to the kinetic energy, grows with doping in the underdoped region and
stops growing at optimal doping.\cite{puch} This also establishes an
important difference between previous $t-J$ model studies and those
reported here.

It has also been observed that in the
direction parallel to the stripes the conduction is much suppressed, as it
can be seen 
in Fig.1b. This is in agreement with previous dynamical studies   
of the model, which showed that the kinetic energy is larger in the
direction perpendicular to the stripes\cite{moham}, as appears also in
Fig.2b. It is also clear that the kinetic energy continuously decreases with
doping in the parallel direction. The corresponding Drude weight is 
denoted with open circles in
Fig.2a.  

\section{Results at Finite Temperature}

The next step is to study the dependence of the optical conductivity
with temperature. Experimental measurements have been
carried out mainly in a narrow range between 0.01 and 0.25 eV, which
corresponds to $\omega<0.50$ in our scale. For
${\rm La_{2-x}Sr_xCuO_4}$ (LSCO) measurements have been reported for 
x=0.13 and 0.14 (underdoped) 
and x=0.22 (overdoped) at temperatures ranging between 10$K$ and 
400$K$,\cite{timusk} while underdoped Bi2212 and optimally
doped Y123 have also been studied.\cite{sb}
\begin{figure}[thbp]
\centerline{\psfig{figure=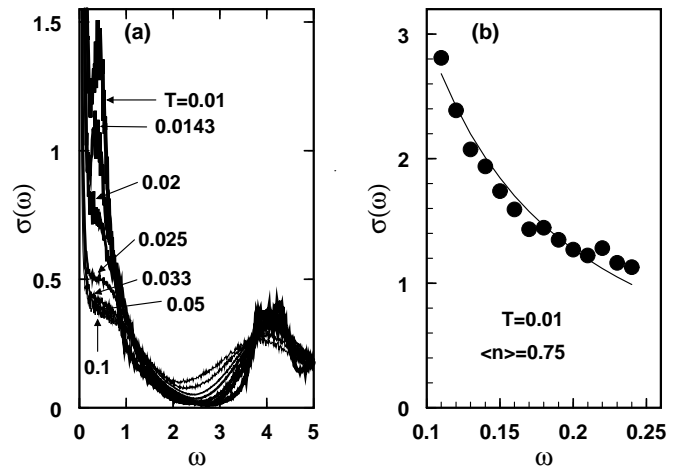,height=6cm}}
\vskip 0.3cm
\caption{(a)The optical conductivity of the Spin-Fermion model versus 
$\omega$ at $\langle n
\rangle=0.75$ in the direction
perpendicular to the stripes and different temperatures on a $12 \times
12$ cluster. The temperatures are indicated; (b) Detail of the curve 
shown in part (a) for
$T=0.01t$ where the data for $0.1<\omega<0.25$ are fitted 
by an $1/\omega$ curve.}
\end{figure}
In the underdoped case, it has been experimentally observed that
as the temperature decreases from 400$K$, spectral weight is transferred
from intermediate toward lower frequencies. 
%the interval 0.025eV-0.087~eV towards lower frequencies. 
This depletion of spectral weight
is associated with the opening of a pseudogap in the density of states
at a temperature $T^*$ estimated to be above 400$K$. In the overdoped and
optimally doped samples,
on the other hand, the depletion of spectral weight is observed
below $T_c$ indicating that $T^* \approx T_c$ in this case.\cite{sb}

In Fig.3a, the optical conductivity versus $\omega$ 
is shown at optimal doping $\langle n \rangle$=0.75 for different
temperatures. Based on the previous studies of 
D-wave pairing correlations, the critical
temperature is $T_c$$\approx$0.025. It can be clearly seen that  spectral
weight is transferred to lower frequencies as the temperature decreases.
The
mid-infrared (MIR) weight increases, as well as the Drude weight, whose
inverse is displayed in Fig.4 as a function of temperature.

The MIR feature is characteristic of  doped cuprates and it is
located at $\omega \approx 0.3$eV.\cite{herr} This is in very good
agreement with our results presented in Fig.1a and 3a where the MIR
feature appears for $\omega \approx 0.5-0.8$, which corresponds to
0.25-0.40~eV for ${\it t}=0.5$eV. 

As mentioned before, it was experimentally observed that at low
frequencies (above $\omega\approx 0.03eV$) the decrease in the
conductivity is closer to $1/\omega$ than $1/\omega^2$, which would have
been expected for free carriers.
The best fit of our data for
$\sigma(\omega)$ in the range $0.1<\omega<0.25$ with a power law
$C/\omega^{\alpha}$ corresponds to $0.9<\alpha<1.3$, in all the
range of temperatures and dopings studied, in excellent
agreement with the experimental results. In Fig.3b we show, as an example, the
$1/\omega$ fit for the data at optimal doping and $T=0.01t$. 
It appears that the mixture of
Drude weight and MIR band at low frequencies 
conspires to produce the $1/\omega$ behavior.

\section{Estimation of the Resistivity}

It has already been mentioned that
the effects of dissipative processes that would create
a finite resistance in a metal have to be added {\it ad-hoc}, by replacing the
delta-functions at $\omega=0$ by Lorentzians, since these processes are not
included in the Hamiltonian of the Spin-Fermion model. Once this is done,
the inverse of the Drude weight should provide information about the dc
resistivity $\rho_{dc}$.\cite{elbio} Transport experiments indicate 
that a linear behavior 
occurs for $T>T^*$, while in the pseudogap region of the underdoped
compounds a faster decrease of
$\rho_{dc}$ is 
found for $T_c<T<T^*$.\cite{timusk,battlog} A power law dependence
$\rho_{dc}\approx T^{1+\delta}$ was observed in the overdoped 
regime with $\delta=0.5$ at 34\% doping in LSCO which is optimally doped
at 15\%.\cite{battlog} 

It is remarkable that many of the above mentioned experimental
characteristics are qualitatively captured by the SFM.
In particular, in Fig.4, a {\it linear} fit of our data at
optimal doping is obtained above $T_c$ (filled circles). 
Notice that, in principle, 
since no dissipative processes are included in the Hamiltonian it is not
possible to determine the onset of superconductivity by studying $1/D$
as a function of the temperature since we cannot distinguish
between a perfect metal and a superconductor by monitoring the Drude
weight in this case. However, we observe a rapid reduction of $1/D$ at
$T_c$ (see Fig.4). This reduction occurs as the D-wave pairing
correlations develop a robust value at long distances. The longest
distances in our clusters is L=6 and as a consequence 
D(6) is also shown in Fig.4 as a
function of the temperature (open circles). 
In this case $T_c=0.025t\approx 150K$, which
is the temperature for which a pseudogap opens in the density of states,
completing an impresive agreement with cuprate's phenomenology.

\begin{figure}[thbp]
\centerline{\psfig{figure=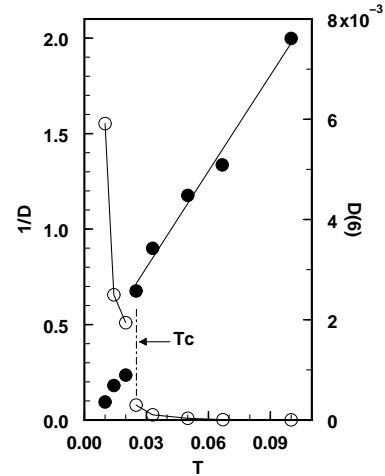,height=6cm}}
\vskip 0.3cm
\caption{The inverse of the Drude weight (filled circles, scale on
the left) 
as a function of the temperature, for the optimal doping 
$\langle n \rangle=0.75$, and for currents perpendicular to the
stripes. The line indicates a linear fit. The open circles indicate the
D-wave pairing correlation at the maximum distance D(6) (scale on the
right). The qualitative agreement with cuprate's experiments is evident.}
\end{figure}

Behavior in qualitative agreement with the experiments is also observed
away from optimal doping. In our underdoped regime, with a
representative density 
$\langle n \rangle=0.875$, the inverse of
the Drude weight and the D(6) pairing correlation are shown as a
function of temperature in Fig.5a. In this case, linear behavior
reducing $T$ is
observed up to $T=0.033t\approx 200K$ when a pseudogap starts developing
in the density of states. Thus, this temperature may be associated to
the experimental $T^*$. Below this temperature, the slope of $1/D$
increases and a further reduction of the resisitivity 
is observed at $T_c=0.017t\approx
100K$ which is the temperature where the D(6) pairing correlations start
to develop. This is in qualitative, although not quantitative, agreement
with the experimental data for ${\rm La_{2-x}Sr_xCuO_4}$.

Data in the overdoped regime, for $\langle n \rangle=0.68$, are
displayed in Fig.5b. As in the previous two cases, a sharp decrease in
$1/D$ is observed at $T_c$ when the D(6) pairing correlations start to
increase. However, above $T_c$, we observe indications of superlinear behavior 
like experimentally observed\cite{battlog}, particularly for the points
at the lower temperatures.

\begin{figure}[thbp]
\centerline{\psfig{figure=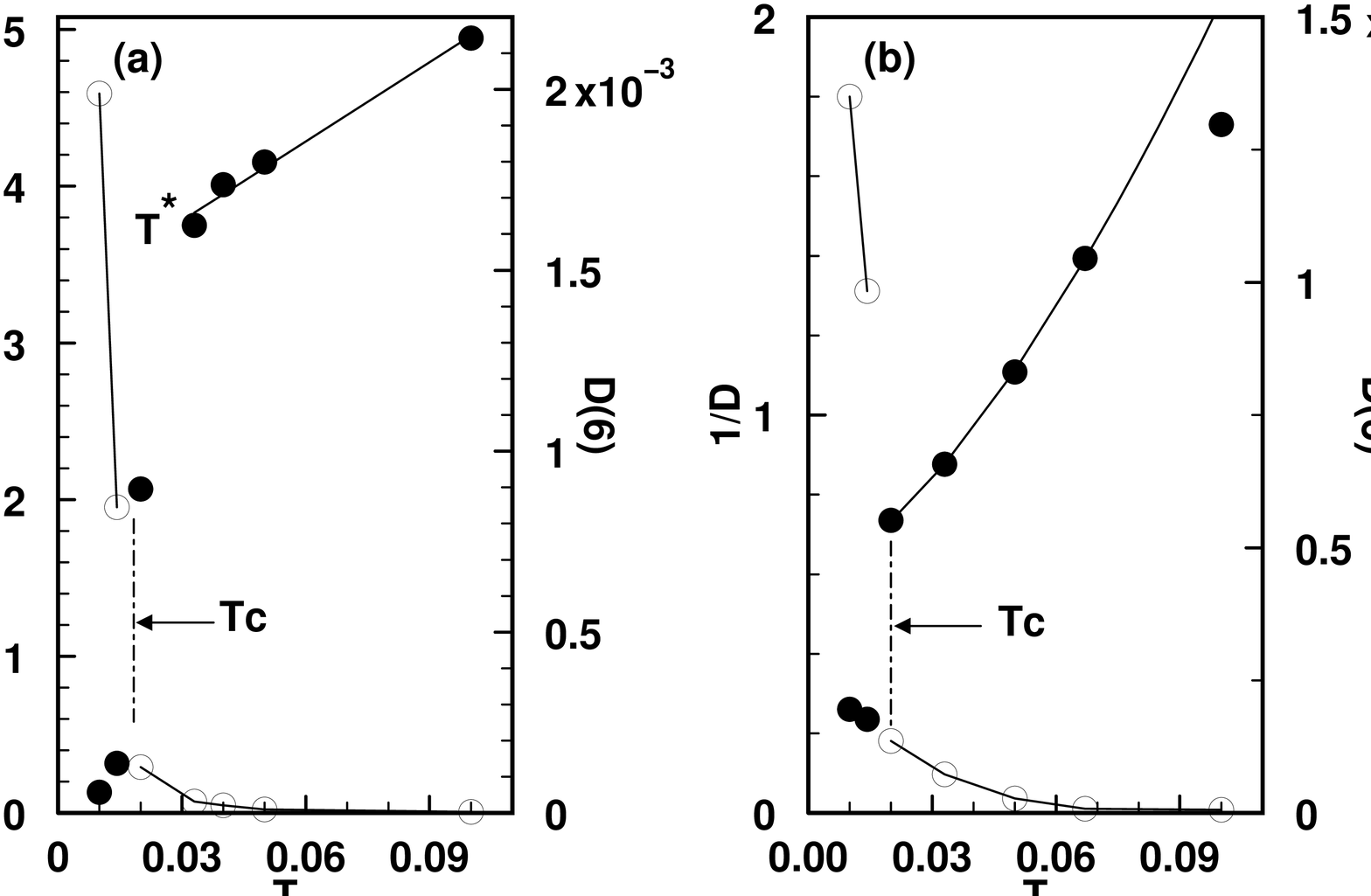,width=8cm}}
\vskip 0.3cm
\caption{(a) Same as Fig.4 but for $\langle n \rangle=0.875$
(underdoped). The continuous line is a linear fit of the $1/D$ points
above $T^*$; (b) Same as Fig.4 but for $\langle n \rangle=0.68$
(overdoped). The continuous line is a $T^{1.5}$ fit of the four first 
$1/D$ points above $T_c$.}
\end{figure}

On the other hand, while the onset of
superconductivity is very obvious in resistivity measurements, it is not
clear that the incoherent part of the 
optical conductivity is sensitive to it,
particularly in the overdoped regime. The transference of spectral
weight from higher to lower frequencies observed as the temperature
decreases has been associated to the opening of the pseudogap rather
than to the onset of superconductivity. No particular feature is
observed at $T_c$ in this case.\cite{timusk}

An interesting characteristic of the SFM results
is that spectral weight is transferred
to lower frequencies when the temperature decreases, as experimentally
observed in the cuprates. This behavior is not obvious. In fact, since
the SFM develops stripes, i.e., charge ordering, at low temperatures
one possible behavior would have been that spectral weight was
transferred toward higher frequencies when the temperature decreases which
is the expected behavior for systems that develop charge density waves.

\section{Conclusions}

Summarizing, the optical conductivity of the SFM appears to have many
features in common with those of the high $T_c$ cuprates. A transference
of spectral weight from high to low frequencies is observed upon doping.
The Drude weight reaches its maximum value at optimal doping and its
inverse, roughly proportional to the resistivity, decreases linearly
with temperature in this regime. Qualitative agreement with the
experimental behavior of the resistivity is also observed in the
underdoped and overdoped cases.
The simplicity of the SFM allows the
study of larger clusters than with the Hubbard and $t-J$ Hamiltonians,
and a much broader range of temperatures can be explored as well. Our
results establish the Spin-Fermion model as a qualitatively realistic
model for cuprates.

We would like to acknowledge useful comments by E. Dagotto, A. Millis
and D. Scalapino. A.M. is supported by NSF under grant DMR-0122523.
Additional support is provided by the National High Magnetic Field Lab 
and MARTECH.

\end{document}